\documentclass[12pt]{iopart}
\usepackage{iopams}  

\usepackage[numbers]{natbib}

\usepackage{url}
\usepackage[scaled]{helvet}
 
\usepackage[T1]{fontenc}
\usepackage{graphicx}
\usepackage{color}

\usepackage{fixltx2e}

\begin{document}

\title[Urban Kaya Relation]{The efficient, the intensive, and the productive: insights from urban Kaya scaling}

\author{Ramana Gudipudi$^1$, Diego Rybski$^1$, Matthias K.\ B.\ L\"udeke$^1$, Bin Zhou$^1$, Zhu Liu$^{2,3,4}$, J\"urgen P.\ Kropp$^{1,5}$}

\address{$^1$ Potsdam Institute for Climate Impact Research, 14473, Potsdam, Germany}
\address{$^2$ John F.\ Kennedy School of Government, Harvard University, Cambridge, Massachusetts 02138, USA}
\address{$^3$ Resnick Sustainability Institute, California Institute of Technology, Pasadena, California 91125, USA}
\address{$^4$ Cambridge Centre for Climate Change Mitigation Research, Department of Land Economy, University of Cambridge, 19 Silver Street, Cambridge CB3 9EP, UK}
\address{$^5$ Dept.\ of Geo- and Environmental Sciences, University of Potsdam, 14476, Potsdam, Germany}


\begin{abstract}
Urban areas play an unprecedented role in potentially mitigating climate change and supporting sustainable development.
In light of the rapid urbanisation in many parts on the globe, it is crucial to understand the relationship between settlement size and CO\textsubscript{2} emission efficiency of cities.
Recent literature on urban scaling properties of emissions as a function of population size have led to contradictory results and more importantly, lacked an in-depth investigation of the essential factors and causes explaining such scaling properties.
Therefore, in analogy to the well-established Kaya Identity, we develop a relation combining the involved exponents.
We demonstrate that application of this \emph{Urban Kaya Relation} will enable a comprehensive understanding about the intrinsic factors determining emission efficiencies in large cities by applying it to a global dataset of 61 cities.
Contrary to traditional urban scaling studies which use Ordinary Least Squares (OLS) regression, 
we show that the Reduced Major Axis (RMA) is necessary when complex relations among scaling exponents are to be investigated.
RMA is given by the geometric mean of the two OLS slopes obtained by interchanging the dependent and independent variable.
We discuss the potential of the Urban Kaya Relation in main-streaming local actions for climate change mitigation.
\end{abstract}

\maketitle

\section{Introduction}
Harbouring more than 50\,\% of the global population \cite{UnitedNations2014}, contemporary cities generate 80\,\% of the Gross Domestic Product (GDP) while consuming approximately 70\,\% of the energy supply and releasing approximately three quarters of global CO\textsubscript{2} emissions \cite{Seto2014}.
Their unprecedented scale and complexity led to the development of a science of cities \cite{Batty2012}. 
Drawing parallels between the allometric scaling in biological systems to that of cities, it has been studied how certain socioeconomic and environmental indicators in cities scale as a function of city size by means of the \emph{urban scaling} approach \cite{BettencourtLHKW2007}.
Since a large fraction of the global population is expected to live in cities by end of this century \cite{Batty2011}, contemporary and future cities will play a pivotal role in global sustainability and climate change mitigation.
Given this strong global urbanisation trend, one of the crucial questions that needs to be addressed is whether large cities are more or less emission efficient in comparison to smaller cities.

The application of urban scaling has triggered copious research in the contemporary science of cities \cite{Samaniego2008,ArbesmanKS2009,AlvesRLM2013,Pan2013,YakuboSK2014,AlvesMLR2015,BettencourtL2016}. 
Urban scaling relates a city indicator (e.g.\ total urban energy consumption) 
with city size (e.g.\ population). 
Assuming power-law correlations, the analysis depicts how these indicators scale with population size and whether large cities are more or less efficient.
A sub-linear scaling (i.e.\ slope $\beta<1$) indicates that large cities consume less, e.g., energy given their size, 
while a unit slope ($\beta\simeq 1$) depicts proportionality,
and a super-linear scaling ($\beta>1$) indicates that large cities consume more energy given their size.

The state-of-the-art research aiming at identifying whether large cities are more energy and emission efficient led to contradictory results and have been largely limited to cities in the developed world. 
E.g., for total CO\textsubscript{2} from cities in the USA, one study reported almost linear scaling \cite{FragkiasLSS2013}, while another one reported super-linear scaling \cite{OliveiraAM2014}. 
A similar study for European cities depicted super-linear scaling \cite{BettencourtL2016}. 
Studies on household electricity consumption in Germany and Spain revealed an almost linear scaling \cite{BettencourtLHKW2007,Horta-Bernus2015}. 
With respect to energy consumed and the subsequent emissions from urban transportation at a household level in the USA, Glaeser \& Kahn \cite{GlaeserK2010} found a sub-linear scaling between population size and gasoline consumption; while another study depicted a super-linear scaling of emissions with population size \cite{LoufB2014SciRep}. 
A similar study done on British cities \cite{Mohajeri2015} found a linear scaling between transport emissions and population size while finding a super-linear relationship between emissions and the total street length. 
Most of the existing studies followed different city definitions and chose different indicators to analyse scaling. 
However, little is known regarding the underlying systematic dynamics that govern these properties.
Therefore, in this paper we develop a framework to investigate the intrinsic factors that determine scaling properties of urban emissions.

We tackle the problem from a different perspective and transfer the idea of the well-established Kaya Identity to urban CO\textsubscript{2} emissions leading to an Urban Kaya Relation. 
Then the scaling of CO\textsubscript{2} emissions with city size can be attributed to the scaling between population, GDP, energy, and emissions. 
To the best of our knowledge, such an attempt to obtain a deeper insight into the scaling of emissions with population using indicators in the Kaya Identity is unprecedented. 
Further, we apply the Urban Kaya Relation to a global dataset of 61 cities.
The objective is to demonstrate its applicability and draw some exploratory empirical conclusions about factors contributing to emission efficiency in large cities globally.
Recent literature has identified that the energy consumption and the subsequent emissions depend on the city type (i.e affluent and mature cities in developed countries versus cities in transition countries with emerging and nascent infrastructure) \cite{CreutzigBBPS2015,Seto2014}. Therefore, we apply the Urban Kaya Relation to these cities separately.

\section{Urban Kaya Relation}
\label{sec:ukr}
The \emph{Kaya Identity} has been proposed to separate global CO\textsubscript{2} emissions into contributions from global population, GDP per capita, energy intensity, and carbon intensity \cite{YamajiMNK1991,Nakicenovic2000,Nakicenovic2004,RaupachMCQCKF2007}.
It relates CO\textsubscript{2} emissions ($C$), population ($P$), GDP ($G$), and energy ($E$) according to
\begin{equation}
C=P \, \frac{G}{P} \, \frac{E}{G} \, \frac{C}{E}
\, .
\label{eq:kaya}
\end{equation}
While the GDP per capita ($G/P$) is a common quantity, the energy intensity ($E/G$) can be understood as the energy necessary to generate GDP, and the carbon intensity ($C/E$) as the efficiency in energy production and consumption (technological).
Equation~(\ref{eq:kaya}) is an identity since it cancels down to $C=C$.

As outlined above, here we are interested in how the urban CO\textsubscript{2} emissions scale with urban population size, i.e.\
\begin{equation}
C \sim P^\phi
\, .
\label{eq:CPphi}
\end{equation}
The value of $\phi$ tells us if large or small cities are more efficient in terms of CO\textsubscript{2} emissions.
We propose that the other quantities also exhibit scaling, i.e.\
\begin{eqnarray}
G & \sim & P^\beta \label{eq:GP}\\
E & \sim & G^\alpha \label{eq:EG}\\
C & \sim & E^\gamma \label{eq:CE} \, .
\end{eqnarray}
Super-linearity of Eq.~(\ref{eq:GP}) with $\beta>1$ is well known in agglomeration economics, see e.g.\ \cite{SveikauskasL1975}, and has recently been confirmed \cite{BettencourtLHKW2007}.
Equation~(\ref{eq:EG}) has been studied on the country scale \cite{BrownBDLDHMNOWZ2011}.
The established power-law relations Eqs.~(\ref{eq:CPphi})-(\ref{eq:EG}) indicate that also Eq.~(\ref{eq:CE}) holds.
In case the power-law form is not empirically supported, Eqs.~(\ref{eq:CPphi})-(\ref{eq:CE}) can still be considered as linear approximations (in log-log space) of potentially more complex functional forms.

In a sense, the exponents $\beta,\alpha,\gamma$ take the role of GDP per capita, energy intensity, and carbon intensity in the original Kaya Identity, Eq.~(\ref{eq:kaya}).
Combining Eq.~(\ref{eq:CPphi})-(\ref{eq:CE}) leads to
\begin{equation}
\phi = \beta\,\alpha\,\gamma \label{eq:ukaya}
\, .
\end{equation}
Thus, in analogy to the original Kaya Identity, Eq.~(\ref{eq:ukaya}) represents an \emph{Urban Kaya Relation} according to which the exponent relating emissions and population is simply given by the product of the other involved exponents.
This permits us to attribute non-linear scaling of emissions with city size [Eq.~(\ref{eq:CPphi})],
to potential urban scaling of GDP with population, energy with GDP, or emissions with energy.
For the sake of completeness, in \ref{sec:k2k3} we also provide another two complementary forms of Kaya Identities and corresponding Urban Kaya Relations.

However, the exponent $\phi$ is usually obtained from data and a linear regression $\ln C = \phi\,\ln P + a$, where $a$ is another fitting parameter. 
Equations~(\ref{eq:CPphi})-(\ref{eq:CE}) represent idealisations and in practice correlations are studied which can come with more or less spread around the regression. Ordinary Least Squares (OLS) might make sense, when dependent and independent variables are clearly defined, e.g.\ in the case of GDP vs.\ population it might be preferable to minimise residuals of GDP. 
Applying Ordinary Least Squares to $C\sim P^\phi$ and $P\sim C^{1/\phi^*}$ generally leads to $\phi\ne\phi^*$ \cite{FluschnikKRZRKR2016,RybskiRWFSK2017} so that also Eq.~(\ref{eq:ukaya}) would not hold (see \ref{ssec:ols}).
In our context, however, dependent and independent variables need to be exchangeable and 
we obtained robust results ($\phi= \phi^*$) by applying \emph{Reduced Major Axis} regression (RMA, see \ref{ssec:rma}).
Therefore, we apply RMA throughout the paper unless specified otherwise. 
In RMA, the slope is given by the geometric mean of the two OLS slopes obtained by interchanging the dependent and independent variable \cite{IsobeFAB1990,BabuF1992}. 
In order to quantify the uncertainty of the estimated exponents, we explore bootstrapping, applying \ 20,000 replications.

\section{Data}
\label{sec:data}
The major pre-requisites while investigating the scaling effects of urban energy consumption and emissions are (a) a consistent definition and demarcation of cities from their hinterlands and (b) a consistent accounting approach to quantify the energy consumption and subsequent emissions \cite{Seto2014}.
The analysis conducted in this paper is limited to 61 global cities, i.e.\ the union of cities for which the 4 quantities are available, i.e.\ (i) total final energy consumption, (ii) CO\textsubscript{2} emissions, (iii) GDP, and (iv) population. 
Although, the data used in this analysis might be inconsistent owing to the challenges mentioned above, we used it 
as a showcase
to demonstrate the applicability of the Urban Kaya Relation. 
The limitations of the data and its implications on the exploratory results are discussed in the Sec.~\ref{sec:disc}.

The population, GDP, and total final energy consumption data used in this study is taken from the Chapter 18 ``Urban Energy Systems'' of the Global Energy Assessment \cite{GrublerBBD2013}. 
This database includes the per capita total final energy consumption of 223 global cities, their respective population and GDP for the year 2005.
The data on emissions is compiled from various sources including city specific reports (provided by organisations such as ICLEI \cite{ICLEI2009}, CDP \cite{CDP}, and C40 cities \cite{C40Cities}) and data which is published in peer reviewed journals \cite{ShanGLL2016}. 

The cities with available data are located in 12 countries.
The GDP per capita of these countries shows two groups. One ranging from 740~USD to 4,700~USD and the other from 26,000~USD to 44,000~USD (year~2005).
These two groups can be considered developing and developed countries and represent the Non-Annex~1 and Annex~1 countries as reported by the United Nations Framework Convention on Climate Change (UNFCCC), respectively.
Amongst the 61 cities used in this analysis 22 cities are from the Annex~I countries 
and 39 cities from Non-Annex~I countries.
The database consists of cities of varying population sizes across 6 continents including 7 mega-cities (with a reported population above 10 million). 
Within countries in Annex~1 regions, 7 cities in the USA, 4 cities in the UK, 2 cities in Germany, Spain, Australia, Italy, France, respectively, and 1 city in Japan are considered in this study. 
With respect to cities in Non-Annex~1 countries 33 cities in China, 2 cities in India, South Africa, and Brazil, respectively, were included.  

On the country scale, CO\textsubscript{2} emissions per capita strongly depend on the development of the considered country, see e.g.\ \cite{CostaRK2011} and references therein. 
Here we pool together cities from many different countries, including from developing countries; as a consequence, the data needs to be normalised prior to the analysis in order to account for baseline emissions, data and other inhomogeneities. 
Therefore, we employ a method that was recently proposed for urban scaling \cite{BettencourtL2016} and normalise the data for each country by the average logarithmic city size ($\langle \ln P \rangle$) and indicator value (e.g.\ $\langle \ln C \rangle$), whereas for each indicator we take the maximum available sample size.

\section{Results}
\begin{figure}
\centering
\includegraphics[width=0.7\textwidth]{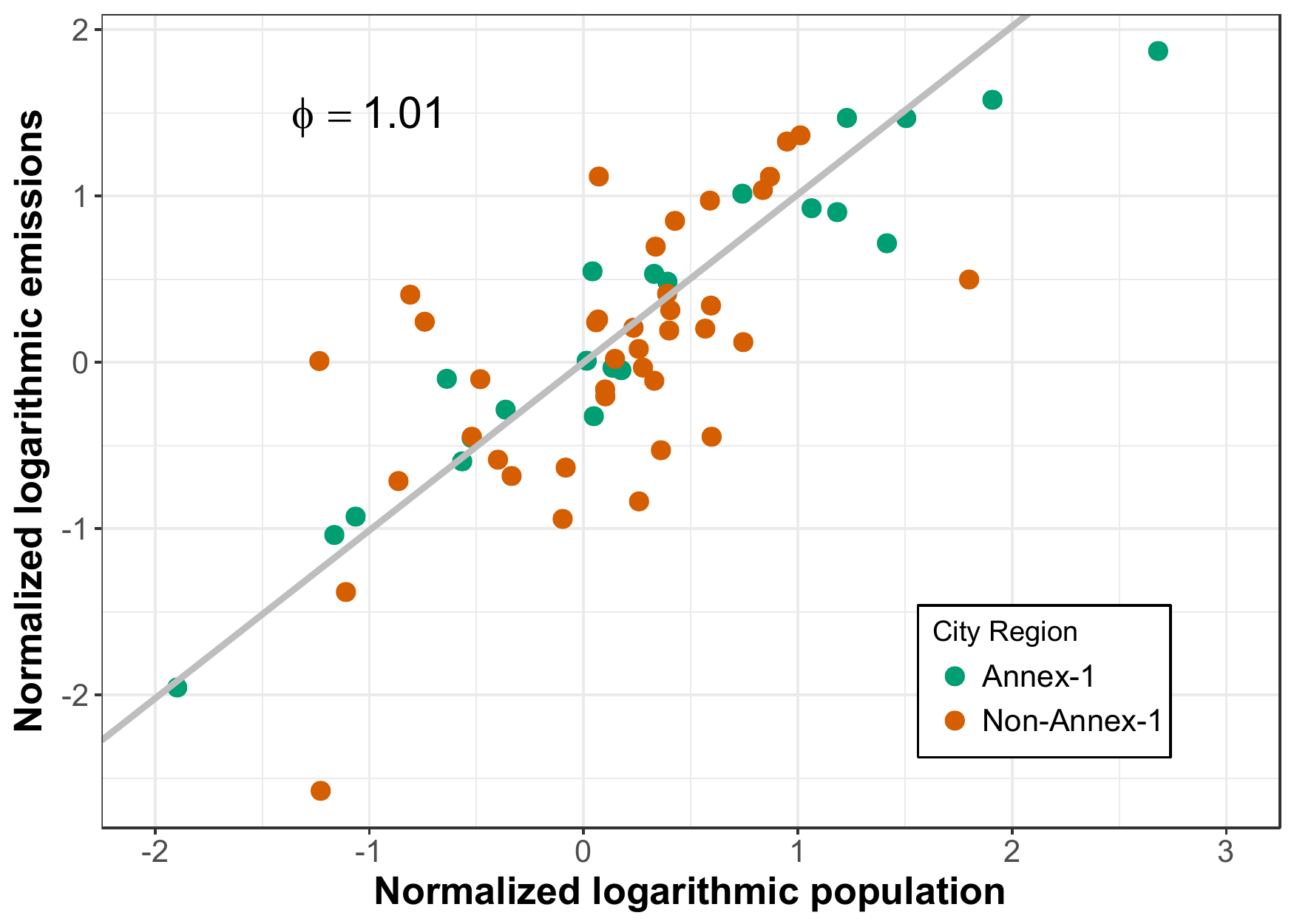}
\caption{Scaling of CO\textsubscript{2} emissions with population size for 61 global cities. The data has been normalised by subtracting average logarithmic values, see Sec.~\ref{sec:data}.
The solid grey line is the regression obtained from the \emph{Reduced Major Axis} (RMA, Sec.~\ref{sec:ukr}) and the slope is $\phi\simeq 1.01$, see Tab.~\ref{tab:01} for details.
}
\label{fig:01}
\end{figure}

We begin by looking at the scaling of emissions with population size for the considered 61 cities. 
The slope of this logarithmic RMA (see Fig.~\ref{fig:01}) is almost equal to one ($\phi=1.01$), however, the pattern of residuals is diverse as also reported in some earlier studies \cite{BettencourtL2016}. 
This result shows that at a global scale large cities are typically not more emission efficient compared to smaller cities.
Further, in Fig.~\ref{fig:01} a distinction between the cities in developed countries (Annex~1) and cities in developing countries (Non-Annex~1) is made.

For comparison, in Tab.~\ref{tab:01} we list the resulting exponents, when we employ OLS to the scaling of the 61 cities. 
Table~\ref{tab:01} also includes the absolute difference between the prediction [Eq.~(\ref{eq:ukaya})] and the measured exponent $\phi$. 
The obtained exponents deviate strongly when OLS is used (instead of RMA).
As discussed at the end of Sec.~\ref{sec:ukr}, we attribute this discrepancy to different regressions when minimising the residuals along any of both axes.
In the case of RMA, plotting $G$ vs.\ $P$ and $P$ vs.\ $G$ leads by definition to the same result.
Thus, we recommend to employ RMA instead of OLS when studying the Urban Kaya Relation. 
Moreover, for OLS it has been shown that whether the estimated exponents are statistically different from 1 depends on the assumptions made \cite{LeitaoMGA2016}.

\begin{figure}
\includegraphics[width=\textwidth]{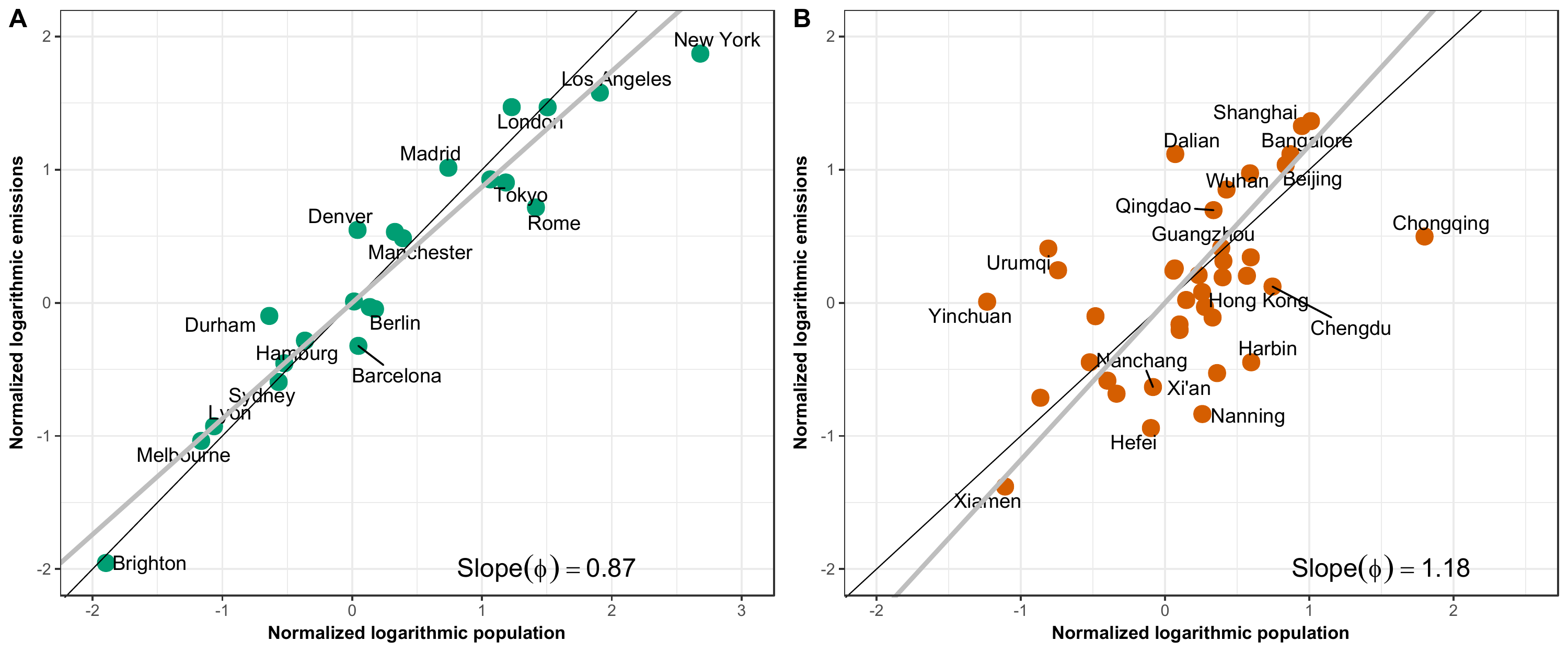}
\caption{Scaling of CO\textsubscript{2} emissions with population for cities in Annex~1 countries (panel~A) and in Non-Annex~1 countries (panel~B). 
Each circle reflects the emission intensity and GDP per capita for a given city in Annex~1 (panel~A) and Non-Annex~1 countries (panel~B), respectively. 
While the slope of the RMA (grey) for Annex~1 countries is found to be sub-linear ($\phi\simeq 0.87$), it is super-linear ($\phi\simeq 1.18$) with respect to cities in Non-Annex~1 countries. 
As in Fig.~\ref{fig:01} the data of both axes has been normalised subtracting average logarithmic values, see Sec.~\ref{sec:data}.
The black line indicates a slope of 1 and is included for comparison.
}
\label{fig:02}
\end{figure}

As a next step, we analysed the scaling properties of emissions with size separately depending on the economic geography of the country (i.e.\ Annex~1 cities vs.\ Non-Annex~1) in which these cities are located. 
In Fig.~\ref{fig:02} we see that the scaling of emissions with the population size indeed has a dependence on the economic geography of the country. 
We found a sub-linear scaling for cities in Annex~1 regions ($\phi=0.87$) and a super-linear scaling for cities in the Non-Annex~1 regions ($\phi=1.18$), see Tab.~\ref{tab:01}. 
In order to test if these slopes are significantly different, we perform bootstrapping and a Kolmogorov-Smirnov (KS) test. 
The KS distance between these bootstrapped samples is 0.83 with a significant P-value ($<2.2\times 10^{-16}$) which confirms that the slopes are not drawn from the same distributions.
The fit appears to be good for cities in Annex~1 regions which are broadly characterised as service sector oriented economies. 
However, in industry dominated Non-Annex~1 cities with widely varying infrastructure and energy intensity of production the goodness of fit appears to be relatively poor. This result shows either that the emissions data from Non-Annex 1 cities is not as accurate, or that population is a good proxy to estimate emissions for cities in Annex 1 regions while there seems to be other factors that influence emissions for cities in Non-Annex~1 countries.

\begin{table}
\caption{Scaling exponents and Urban Kaya Relation. The Table lists the various estimated exponents and the last column shows how well the Urban Kaya Relation performs. 
The exponents are listed for all cities, cities in Annex~1 countries, and cities in Non-Annex~1 countries (see Sec.~\ref{sec:data}).
All exponents have been obtained from RMA (see Sec.~\ref{sec:ukr}) except for the last row, where OLS has been applied for comparison. 
The square brackets give 95\,\% confidence intervals from bootstrapping (20,000~replications).
Inspired by the notation used in \cite{LeitaoMGA2016} we put the following symbols. $\nearrow$, at least 66.6\,\% of the replications lead to exponents larger than $1$; $\rightarrow$, 33.3\,\% to 66.6\,\% of the estimates are larger than $1$, and $\searrow$, less than 33.3\,\% are larger than $1$.
While Eq.~(\ref{eq:ukaya}) works exactly for RMA (see Appendix C.2), for OLS the estimated exponents are incompatible (last row).
}
    \begin{tabular}{p{2cm}|p{2cm}|p{2cm}|p{2cm}|p{2cm}|p{3cm}}
    \hline
Exponent: & $\phi$ & $\beta$ & $\alpha$ & $\gamma$ & $|\phi-\beta\,\alpha\,\gamma|$ \\
Equation: & Eq.~(\ref{eq:CPphi}) & Eq.~(\ref{eq:GP}) & Eq.~(\ref{eq:EG}) & Eq.~(\ref{eq:CE}) & Eq.~(\ref{eq:ukaya}) \\
Scaling of: & Emissions with population & GDP with population & Energy with GDP & Emissions with Energy &  \\
\hline
    All Cities & 1.01 $\rightarrow$ [0.87,1.18]& 1.13 $\nearrow$ [1.04,1.24] & 0.92 $\searrow$ [0.78,1.06] & 0.97 $\rightarrow$ [0.82,1.14]& 0.00 \\
    \hline
    Annex~1 & 0.87 $\searrow$ [0.59,0.95] & 1.03 $\nearrow$ [1.00,1.15] & 0.99 $\rightarrow$ [0.60,1.63] & 0.85 $\searrow$ [0.41,1.21] & 0.00 \\
    \hline
    Non-Annex~1 & 1.18 $\nearrow$ [0.90,1.49] & 1.27 $\nearrow$ [1.04,1.55] & 0.83 $\searrow$ [0.67,1.00] & 1.11 $\nearrow$ [0.93,1.29]  & 0.00 \\
    \hline
    All Cities (OLS)  & 0.80 & 0.98 & 0.64 & 0.67 & 0.38\\
    \hline
\end{tabular}
  \label{tab:01}
\end{table}

We looked at scaling of each of the indicators in the Urban Kaya Relation, namely the scaling of GDP with population ($G/P$) Eq.~(\ref{eq:GP}), scaling of energy intensity ($E/G$) Eq.~(\ref{eq:EG}), and carbon intensity ($C/E$) Eq.~(\ref{eq:CE}). 
Table~\ref{tab:01} lists the exponents for each of these relations. 
From a global perspective, our results suggest that the almost linear scaling of emissions with population size could be attributed to the almost linear scaling of carbon intensity and the trade-off between scaling of GDP with population and the scaling of energy intensity (i.e.\ they compensate each other). 

In the case of cities in Annex~1 countries, our results show that the large cities typically have lower emissions per capita compared to smaller cities because of the sub-linear scaling of the carbon intensity (Tab.~\ref{tab:01}).
This might be attributed to the carbon intensity of the electricity generation supply mix, vehicle fuel economy, and the quality of public transit in these cities \cite{KennedySGHHHPPRM2009}.
We found an approximately linear scaling of GDP with population. 
Our result shows that doubling the GDP in these cities will lead to an almost similar increase in energy consumption. 
Such a linear scaling might be largely attributed to the consumption patterns and infrastructure lock-in behaviour in largely service based economies \cite{Satterthwaite2009,CreutzigBBPS2015}.   

We further checked if the sub-linear scaling of emissions with population for cities in Annex~1 countries could be attributed to a possible sub-linear scaling with respect to their total final energy consumption Eq.~(\ref{eq:EP}). 
Even in a completely decarbonised world, the question of energy efficiency will persist. 
Our results suggest that large cities in Annex~1 countries are not much more energy efficient with respect to their population ($\delta\simeq 1.04$, see \ref{sec:k2k3}) compared to smaller cities. 
This result indicates that although the per capita energy consumption in large cities is similar to that of smaller cities, it is the better technologies employed in larger cities that typically make their per capita emissions lower than in smaller cities. 

With respect to cities in Non-Annex~1 countries, our results suggest that the super-linear scaling of emissions with population is due to two factors: (1) super-linear scaling of GDP with population and (2) super-linear scaling of carbon intensity. 
However, we found that doubling the GDP in these cities will lead to a less than double increase in energy consumption. 
This might be attributed to the prevalence of energy poverty in these cities \cite{Satterthwaite2009}. 
Large cities in Non-Annex~1 countries benefit economically (more GDP) from the urban poor who consume less energy and have limited access to electricity. 
Therefore, large cities in this region are more energy efficient compared to smaller ones.

\section{Discussion \& Conclusions}
\label{sec:disc}
In summary, the achievements of this work are threefold.
(i) In analogy to the Kaya Identity -- which in the climate change community represents a well know specification of the IPAT approach (see \ref{sec:ipat}) -- we set out a framework to assess why urban CO\textsubscript{2} emissions scale super- or sub-linearly with city size [Eq.~(\ref{eq:CPphi})]. We derive the \emph{Urban Kaya Relation} $\phi=\beta\,\alpha\,\gamma$.
(ii) We show that Ordinary Least Squares (OLS) lead to erroneous results and propose to use \emph{Reduced Major Axis} (RMA) regression.
(iii) As a proof of concept we apply the Kaya framework to the available data.
In the first place, the proposed Kaya relation can be used to see from which (in)\-efficiency $\phi\ne 1$ is stemming from. In the second place, it can serve as consistency check, i.e.\ the product of exponents must be correct.

It is crucial to establish foundations in the form of such a framework to understand the guiding factors that govern scaling properties since urban areas are often identified as the focal spatial units for improving energy efficiency and climate change mitigation \cite{DodmanD2009,HoornwegSG2011}.
Urban energy consumption and subsequent emissions is an outcome of urbanites' affluence and their consumption patterns \cite{SatterthwaiteD2008}. 
Nevertheless, it is important to investigate whether the infrastructural efficiency of large cities will be manifested as emission efficiency gains.
An in-depth investigation about the demographic, economic and technological drivers of urban emissions is necessary to identify the key entry points for mitigation actions at a city scale.
By means of an exploratory analysis we demonstrate that the Urban Kaya Relation can be used to address this issue adequately by attributing the scaling properties of emissions to the scaling of GDP with population (affluence), energy intensity (economic geography) and emission intensity (technology).

The data used in this study has three major constraints. 
Firstly, the sources for emission data is different from the source of energy consumption and GDP. 
Therefore, the urban extents and the population size might vary for few cities. 
Studies have shown that such city definitions will influence the scaling properties \cite{ArcauteHFYJB2014,Cottineau2015}. 
Secondly, since the data is from multiple sources, the accounting approach varied in most of cities. 
Inclusion (or exclusion) of emissions embedded in electricity generation will have a significant impact on energy and emissions attributed to building sector. 
Thirdly, the sectoral emissions and type of fuels used varies from one city to another. 
The energy data from the GEA \cite{GrublerBBD2013} study includes the total final energy consumption but excludes traditional biomass for few cities in developing regions. 
Data on emissions attributed to the industrial sector is inconsistent as it excludes industrial electricity and fuel usage in some cities. 
In the view of aforementioned data limitations, the broader conclusions drawn in this paper should be treated cautiously since the empirical part is rather intended as a proof of concept.

These data constraints also highlight the data needs.
Therefore, we acknowledge the ongoing efforts to develop a consistent emission framework by various international organisations\footnote[7]{http://www.ghgprotocol.org/}
and make an appeal that such efforts should disclose data on energy consumption and GDP along with sectoral emissions and population.
Application of the Urban Kaya Relation on such a dataset will enable researchers to identify locally appropriate mitigation actions.

The urban scaling approach can be attributed to city functionality as a (short-term) spatial equilibrium of the interplay between the density dependent socioeconomic interactions and transportation costs \cite{BettencourtL2016}. 
The universality of the scaling properties of socioeconomic (super-linear) and infrastructural components (sub-linear) in cities can be explained as an outcome of the increased interaction between citizens at a microscopic scale \cite{Schlapfer2014} and it's combination with the fractal properties of cities \cite{Ribeiro2017}.
However, unlike socioeconomic and infrastructure components, urban emissions are a result of various economic activities (e.g.\ location of industries). 
Such processes consume resources and energy which are 
often originated beyond the urban boundaries but have an implication on emissions from electricity consumption. 
In such cases, population alone might not be a good predictor for emissions.
This could be the explanation for the relatively poor fitting of the emission scaling properties of cities in Non-Annex~1 regions in comparison to cities in Annex~1 regions.

Our exploratory results show that large cities in Annex~1 countries have lower emissions compared to smaller cities. 
This result suggests the usage of better technologies in energy generation/consumption and efficient modes of transportation.
From a climate change mitigation point of view, the key challenge in these cities is to further decrease their energy and carbon intensity while ensuring economic stability. 
According to our exploratory results larger cities in emerging countries such as China, India, and Brazil typically have more per capita emissions compared to smaller cities. From one point of view, it may be good news that large cities in these regions are not emission efficient since much of the urbanisation in these regions is going to happen in small and medium size cities \cite{DepartmentofEconomicandSocialAffairs2014}.
Thus, despite being exploratory, our findings corroborate the results of previous studies which showed the significance of affluence on emissions \cite{Martinez-Zarzoso2011} and the influence of economic geography on the scaling properties of emissions with population \cite{RybskiRWFSK2017}. Further support comes from a recent study, where a methodology other than urban scaling has been applied and completely different data has been used \cite{ZhouGLBKR2017}.

Last but not least, we need to mention the role of urban population \emph{density} as an important factor in determining the energy consumption and subsequent emissions. 
On the one hand, it has been shown that urban CO\textsubscript{2} emissions from transport energy per capita decrease with population density \cite{Gudipudi2016,KennedySGHHHPPRM2009,CreutzigBBPS2015}. 
On the other hand, there is a theoretical connection between urban indicators, population, and area scalings \cite{RybskiD2016,RybskiRWFSK2017}.
Combining density with the Urban Kaya Relation introduces further complexity which we leave to be addressed by future research.

\vspace{1cm}

\section*{Data accessibility}
All data used are publicly available, sources are referenced in Sec.~\ref{sec:data}.

\section*{Authors' contributions}
RG, DR, MKBL, BZ, JPK designed the study;\\
ZL, RG collected the data;\\
RG carried out the data analysis;\\
RG, DR, MKBL, BZ validated the results;\\
RG prepared the Figures;\\
RG, DR, MKBL, JPK wrote and revised the manuscript.

\section*{Competing interests}
We declare we have no competing interests.

\section*{Research Ethics}
No experiments were carried out during this study and no prior ethical assessment was required.

\section*{Funding}
The research leading to these results has received funding from the European Community's Seventh Framework Programme under Grant Agreement No.~308497 (Project RAMSES).
Zhu Liu acknowledges support by the Green Talents Program held by the German Federal Ministry of Education and Research (BMBF).

\section*{Acknowledgment}
We thank M. Barthelemy, H.\,V.~Ribeiro, and L. Costa for useful discussions.
This work emerged from ideas discussed at the symposium \emph{Cities as Complex Systems} (Hanover, July 13th-15th, 2016) which was generously funded by VolkswagenFoundation.

\appendix

\section{Kaya~II and~III}
\label{sec:k2k3}
It needs to be mentioned that there are another two identities complementary to the original Kaya Identity, Eq.~(\ref{eq:kaya}), namely
\begin{eqnarray}
C &=& P \, \frac{E}{P} \, \frac{G}{E} \, \frac{C}{G} \label{eq:k2}\\
C &=& G \, \frac{P}{G} \, \frac{E}{P} \, \frac{C}{E} \label{eq:k3} \, ,
\end{eqnarray}
or variations.
We propose to denote Eqs.~(\ref{eq:kaya}), (\ref{eq:k2}), and (\ref{eq:k3}), ``Kaya~I'', ``Kaya~II'', and ``Kaya~III', respectively.
The identities Kaya~II and~III involve two intensities which do not appear in Kaya~I, namely $E/P$ and $C/G$, i.e.\ energy per capita and carbon per GDP, respectively.
In the urban scaling picture these take the form
\begin{eqnarray}
E & \sim & P^\delta \label{eq:EP} \\
C & \sim & G^\eta \label{eq:CG} \nonumber \, .
\end{eqnarray}
The relations corresponding to Eqs.~(\ref{eq:k2}) and~(\ref{eq:k3}) are
\begin{eqnarray*}
\phi & = & \frac{\delta\,\eta}{\alpha} \label{eq:ukaya2} \\
\eta & = & \frac{\delta\,\gamma}{\beta} \label{eq:ukaya3} \, .
\end{eqnarray*}
Other combinations of $C$, $P$, $G$, or $E$ involve only two components each.

\section{Different Linear Regression Slopes}
\label{sec:techniques}

\subsection{Ordinary Least Squares (OLS)}
\label{ssec:ols}
If we consider $c_i=\ln C_i$ and $p_i=\ln P_i$ with standard deviations $\sigma_c$ and $\sigma_p$, respectively, then the slope according to $\textrm{OLS}(c|p)$ is analytically given by
\begin{equation}
\phi=\rho_{c,p}\frac{\sigma_c}{\sigma_p}
\label{eq:phiols}
\end{equation}
where $\rho_{c,p}$ is the correlation coefficient \cite{IsobeFAB1990,BabuF1992}.
Accordingly, for Eq.~(\ref{eq:ukaya}) we obtain
\begin{eqnarray}
\beta\,\alpha\,\gamma &=& 
\rho_{g,p}\frac{\sigma_g}{\sigma_p}\,
\rho_{e,g}\frac{\sigma_e}{\sigma_g}\,
\rho_{c,e}\frac{\sigma_c}{\sigma_e}\\
 &=& \rho_{g,p}\rho_{e,g}\rho_{c,e}\frac{\sigma_c}{\sigma_p}
\, .
\end{eqnarray}
Comparison with Eq.~(\ref{eq:phiols}) leads to $\rho_{c,p}=\rho_{g,p}\rho_{e,g}\rho_{c,e}$, which is \emph{not true} in general.

\subsection{Reduced Major Axis (RMA)}
\label{ssec:rma}
The Reduced Major Axis (RMA) is given by the geometric mean of the two OLS slopes, i.e.\ minimizing the sum of squares of vertical residuals, $\textrm{OLS}(y|x)$, or horizontal residuals, $\textrm{OLS}(x|y)$, respectively \cite{IsobeFAB1990,BabuF1992}.
Then the slope according to RMA is analytically given by
\begin{equation}
\phi=\textrm{sign}(\rho_{c,p})\frac{\sigma_c}{\sigma_p}
\label{eq:phirma}
\, .
\end{equation}
Since in our case the correlations are always positive, we omit $\textrm{sign}(\rho)$ from now on.
Then, for Eq.~(\ref{eq:ukaya}) we obtain
\begin{eqnarray}
\beta\,\alpha\,\gamma &=& 
\frac{\sigma_g}{\sigma_p}\,
\frac{\sigma_e}{\sigma_g}\,
\frac{\sigma_c}{\sigma_e}\\
 &=& \frac{\sigma_c}{\sigma_p}
\, ,
\end{eqnarray}
which is consistent with Eq.~(\ref{eq:phirma}).
One can see that the result is independent from any correlation coefficient.

Please note, in a previous version of our manuscript we used Orthogonal Regression (also known as Total Least Squares, TLS).
Since the corresponding analytical expression for the slope is much more complex, here we employ the simpler RMA.

\section{IPAT concept}
\label{sec:ipat}
The original Kaya Identity is a specific version of the IPAT concept, which stands for
\begin{equation}
I = P\, A\, T
\, ,
\end{equation}
where the quantities are impact ($I$), population ($P$), affluence ($A$), technology ($T$), see e.g.\ \cite{ChertowM2000} and references therein.
A stochastic IPAT version introduced in \cite{DietzR1997} is in our notation given by
\begin{equation}
C=a\, P^b\, G^c \, E^d
\label{eq:ipatdietz}
\, ,
\end{equation}
where $a$\dots{}$d$ are parameters.

While Eq.~(\ref{eq:ipatdietz}) is a higher-dimensional extension of Eq.~(\ref{eq:CPphi}), with the goal of better predicting $C=\textrm{f}(P,G,E)$, our approach is still based on Eq.~(\ref{eq:CPphi}) but aims at circumscribing it employing the other scaling relations Eq.~(\ref{eq:GP})-(\ref{eq:CE}).

\bibliographystyle{iopart-num}

\providecommand{\newblock}{}

\end{document}